\documentclass[conference]{IEEEtran}

\usepackage{amsmath,amsfonts}
\usepackage{graphicx,subfigure}

\ifCLASSINFOpdf
\else
\fi

\begin{document}

\title{Monaural Audio Speaker Separation Using \\ Source-Contrastive Estimation}

\author{\IEEEauthorblockN{Cory Stephenson, Patrick Callier, Abhinav Ganesh, and Karl Ni}
\IEEEauthorblockA{Lab41, In-Q-Tel Laboratories\\
Menlo Park, CA 94025\\
Email: lab41@iqt.org}
}

\maketitle

\begin{abstract}
We propose an algorithm to separate simultaneously speaking persons from each other, the ``cocktail party problem'', using a single microphone. Our approach involves a deep recurrent neural networks regression to a vector space that is descriptive of independent speakers. Such a vector space can embed empirically determined speaker characteristics and is optimized by distinguishing between speaker masks. We call this technique source-contrastive estimation. The methodology is inspired by negative sampling, which has seen success in natural language processing, where an embedding is learned by correlating and decorrelating a given input vector with output weights. Although the matrix determined by the output weights is dependent on a set of known speakers, we only use the input vectors during inference. Doing so will ensure that source separation is explicitly speaker-independent. Our approach is similar to recent deep neural network clustering and permutation-invariant training research; we use weighted spectral features and masks to augment individual speaker frequencies while filtering out other speakers. We avoid, however, the severe computational burden of other approaches with our technique.
Furthermore, by training a vector space rather than combinations of different speakers or differences thereof, we avoid the so-called permutation problem during training. Our algorithm offers an intuitive, computationally efficient response to the cocktail party problem, and most importantly boasts better empirical performance than other current techniques.
\end{abstract} 

\IEEEpeerreviewmaketitle

\section{Introduction}
\label{introduction}
Source separation and signal denoising have been a problem in multiple media for over a century with applications ranging from acoustic speech processing to underwater vessel tracking. For each application, techniques addressing the problem have extended to array processing~\cite{bss-beamforming} (to include work in domains like SONAR and RADAR), adaptive signal processing, and component analysis (matrix factorization and probabilistic modeling). The problem setup in the majority of these instances assume that it is well-posed, where information theoretical guarantees can be placed on each source based on their temporal and spatial origins relative to the sensors.

Unfortunately, the same guarantees cannot be extended to the monaural case~\cite{monaural-ica}, where audio is recorded from a single location. In the previous scenarios, assumptions are explicitly made that specify a large degree of control over environment or listening devices, usually in the form of multi-microphone systems or specially designed radiation patterns. The more common scenario, monaural audio, comes about because audio is readily recordable with the proliferation of portable devices (e.g., cellphones). In such scenarios, it is seldom the case that there is enough spatial and temporal information to isolate and denoise individual sources.

In response, related work in the last decade has adapted familiar matrix factorization approaches to spectral features derived from the magnitude response~\cite{monaural-nmf,isik16,monaural-ica,KandT,schmidt-nmf}, though it has quickly become apparent that higher model complexity is required. The necessary model complexity can be modeled with a priori knowledge that is empirically derived from training on data, which set the stage for machine learning for signal processing practices. As neural network approaches have re-entered the mainstream machine learning community, papers in the past couple of years~\cite{dong-permutation,deepkaraoke} have attempted to adapt them to the speech separation problem. 
It is with these methods that we have found that recurrent neural networks for speech processing have shown the most promise in modeling acoustic time series~\cite{hershey2016deep}.

Developing cost functions for neural networks to separate speakers is challenging because traditional machine learning-based approaches are typically rooted in the classification problem. In contrast, speaker separation should be usable when speakers are out of set (i.e., not in the training set). Secondly, label ambiguity is a major issue during training because the actual order separated signals is arbitrary (i.e., whether to call a participant speaker 1 or speaker 2), 
the so-called permutation problem on which many works~\cite{dong-permutation,pit-ref-1} have concentrated their efforts. Their solution is to consider every possible pair of speakers during training, a costly approach. Alternatively, according to our analysis, vector space projection~\cite{hershey2016deep} has shown the best separation to date by minimizing the difference in correlation of a time-frequency bin to all other time-frequency bins. While effective, an inherent inefficiency is exposed in the underlying algorithm: speaker attributes are optimized by how they relate to each other and not by what makes them distinct. Moreover, auto-correlation of time-frequency bins is not only unnecessary, but detrimental to performance. 


Using the referenced works as a starting point, we propose an algorithm to \emph{directly} optimize a vector space that isolates specific speaker characteristics. We borrow concepts from negative sampling based approaches from natural language processing~\cite{mikolov}, and instead train an embedding where speakers are explicitly contrasted with each other. The conjecture is that such an intuitive approach will provide better discrimination where the cost function appropriately models our goal. 
We call the proposed algorithm source-contrastive estimation, as distinguished from noise-contrastive estimation. Our vector space is speaker independent and performs significantly better than state of the art.

The remainder of this paper describes our approach to source separation of finding optimal vector spaces using source-contrastive estimation. We consider and review state of the art in monaural source separation in Sec.~\ref{related}. The approach is then described in Sec.~\ref{approach} with implementation details in Sec.~\ref{implementation}. Experimental results are shown in Sec.~\ref{results}, which is followed by a summarization and discussion of future work

\section{Related Work}
\label{related}
Before the latest turn to deep learning, principal approaches to monaural source separation centered on matrix factorization techniques~\cite{monaural-ica}. Among the more advanced of such algorithms, non-negative matrix factorization (NMF)~\cite{monaural-nmf} is a commonly cited baseline. Attempts to address monaural and underdetermined separation problems in the matrix factorization paradigm either focus on model variance~\cite{schmidt-nmf} or augmenting the number of sensor channels and applying ICA~\cite{KandT}. 



Speech source separation shares commonalities with research in denoising and speech enhancement, and has even closer ties to the vocals-music separation problem~\cite{deepkaraoke}. In all such problems, the categories of signal being separated are qualitatively distinct; aiming to separate out vocals as vocals, instrumentals as instrumentals, and so on. Source separation is more complex, because optimal solutions are identifiable only up to a permutation---it doesn't matter, in other words, whether speaker A ends up on track 1 or track 2, as long as speaker B ends up on the other one. 
Several attempts have been made to resolve this so-called permutation problem, though they come at the price of additional computational complexity. In particular,~\cite{dong-permutation} looks at all possible assignments of speakers to reconstructed outputs, selecting the minimal loss across such assignments. 
In this way,~\cite{dong-permutation}, like most prior work in the field, treat the separation problem as a matter of class-based segmentation.

Deep clustering (DC)~\cite{isik16} not only addresses the permutation problem but is one of the first "partition-based" segmentation models in the source separation field. DC constructs a time-frequency embedding space trained on speaker difference. At inference time, DC can separate mixtures of theoretically any number of speakers, regardless of the number and identity of speakers in the training data. During training, the learning objective of deep clustering encourages the embedding model to generate similar vectors for each time-frequency bin associated with a particular speaker. This necessitates comparing each time-frequency bin to each other time-frequency bin.

DC is related to another approach, deep attractor networks (DA)~\cite{attractor}. DA also projects every time-frequency bin into a low-dimensional embedding space, but its objective is to pull each time-frequency embedding vector for a given speaker toward each other by using attractor vectors in the embedding space.  Separation masks are calculated according to the distance of each embedding to the attractors for the various sources. 
This leads to well-separated vector spaces after training and an elegant method for calculating masks, but selecting attractors for unknown sources during inference relies on somewhat ad hoc measures.

PIT, DC and DA have made significant progress toward solving the source separation problem. Two observations about their contributions motivate the present work. For DC in particular, the pairwise comparison of time-frequency bins to each other is needlessly cumbersome. DA sidesteps this by only comparing each T-F embedding to the $M$ attractor vectors, where $M$ is the number of sources being considered. This greatly reduces the number of comparisons being made, but leads to our second observation: during training for DA, source-specific attractors are induced at the centroids of each source's embedding space---meaning attractor locations depend on properties of the example and of the source, and can change positions arbitrarily. Embedding techniques like word2vec~\cite{mikolov}, in contrast, optimize embeddings of their \textit{targets} in tandem with the embeddings of the inputs. If the goal during training is to maximize similarity to a representation of the target source in the embedding space, a joint optimization approach such as this might offer a more direct solution. At inference time, target-specific embedding models can be discarded and the input embeddings will still retain the desirable properties they acquired during training. Moreover, such an approach could obviate the need to find attractors at inference time, overcoming a major limitation of DA.






\section{Approach}
\label{approach}
Our approach to monaural source separation operates on the assumption that linearly mixed speech $x(t)$ can be well-separated into individual speakers $s_i(t)$. For a given speaker $i$ in a speaker mix, this is most often done by masking the magnitude response, filtering out information from time-frequency bins in the short-time Fourier transform (STFT), $X(t,f)$, that do not belong to him/her, while passing those time-frequency bins that do. 

Typically, the predicted mask $Y_{t,f}^{(i)}$ for the $i^{th}$ speaker is implemented as either a ratio or in our case, a binary mask. We let $Y_i \in \{-1, 1\}^{M \times TF}$, where $M \leq C$, $C$ being the total number of speakers in our training set and $M$ being the number to be mixed. To set our masks, if $i^{th}$ speaker is the loudest in that time frequency bin, we set $Y_{t,f}^{(i)}=1$, and $Y_{t,f}^{(i)} = -1$ otherwise.

\begin{figure}[h]
\centering
\includegraphics[width=0.45\textwidth]{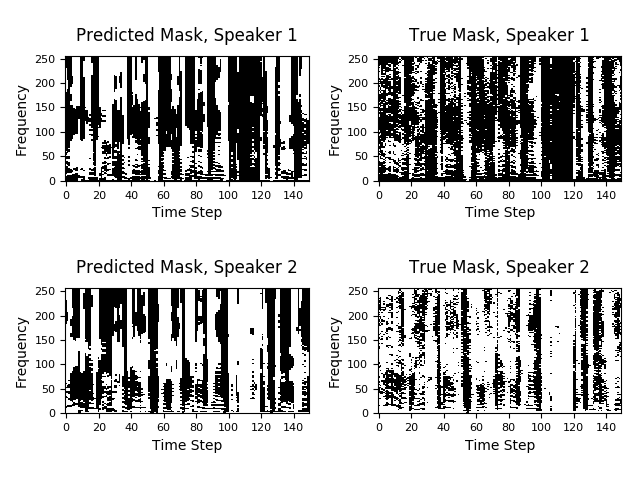}
\caption{Example of predicted (product of using a sigmoid as a nonlinearity) and true binary masks}
\label{algorithm}
\end{figure}


Similar to natural language processing embedding techniques like \emph{word2vec}~\cite{mikolov}, a given word embedding can represent specific words. Instead of a word embedding, we use a speaker embedding, similarly optimized via two vector spaces. The first vector space is an input embedding that \emph{implicitly} defines a speaker, and it is not associated with anyone in particular. We also have an output embedding that \emph{explicitly} trains to a corpus of known speakers. Then, when performing inference to input vector space, it is possible to generalize to any possible speaker by clustering our neural network outputs. In our notation, the input and output vector spaces for a given sample are implemented as tensors with an embedding space of $E$, labeled as $V_i(t,f)$ and $V_o$, respectively. The columns of either tensor have $E$ dimensions (hidden units) and denote the vectors associated with a speaker's likeness. The entire process for training and testing is shown in Fig.~\ref{algorithm}.

To train and generate our embeddings, we use a recurrent neural network regression to $V_i$. To compare to~\cite{hershey2016deep}, we use a total of two BLSTM layers, and we have a dense layer that is convolved over the output 2D vector produced by the final BLSTM. That is where the similarities end, however.

Let our loss for every time frequency bin for sample $b$ be denoted as $\mathcal{L}^{(b)}_{t,f}$. Then,
\begin{equation}
\label{eqn:cost}
\mathcal{L}^{(b)}_{t,f}(\mathbf{v}_i, \mathbf{v}_o) = \frac{-1}{M} \sum_{s \in S_b} \log \sigma \left( Y^{(s,b)}_{t,f} \cdot \textbf{v}^{(b)}_i(t,f)^T \textbf{v}^{(s)}_o \right)
\end{equation}

Here, $S_b$ is the set of speakers sampled for mix $b$, and $s$ is a single speaker from the subset. The total loss for the batch of size $B$ over all frequencies and time is thus,

\begin{equation}
    \mathcal{L}(\mathbf{v}_i, \mathbf{v}_o) = \frac{1}{B} \sum_b \sum_{(t,f)} \mathcal{L}^{(b)}_{t,f}
\end{equation}

Intuitively, the output of the neural network at time $(t,f)$ is $\textbf{v}_i(t,f)$. and the output vector $\textbf{v}^{(s)}_o$ is an embedding for speaker $s$ at frequency $f$. Say that speaker $1$ is louder than speaker $2$ at time frequency bin $(t,f)$ for sample $b$. Then we would ideally like the correlation between the embedding produced by our neural network $\textbf{v}_i$ and the vector for speaker 1 to be high. That is to say, we would like $\sigma (\textbf{v}_i^T \textbf{v}^{(1)} ) \rightarrow 1$. Simultaneously, the correlation between $\textbf{v}_i$ and the vector for speaker 2 should be low, since these two vectors should be anti-correlated if they are sufficiently different. That is to say, we would like $\sigma(\textbf{v}_i^T \textbf{v}^{(2)}) \rightarrow 0$.  Mathematically speaking, we are pulling our embedding towards our speaker vector $\textbf{v}^{(1)}_o$ and pushing it away from non-speaker vectors $\textbf{v}^{(2)}_o$. Which speakers to attribute appropriate correlation/anti-correlation to is determined by the label $Y$, which will be $+1$ in the former case and $-1$ in the latter. It is important to note that we can save on both computation and accuracy by optimizing only those speakers that are in $S_b$, which in our case (we train against two speaker mixes) will have two elements.

\begin{figure*}[h]
\centering
\subfigure[Training]{
\includegraphics[width=0.85\textwidth]{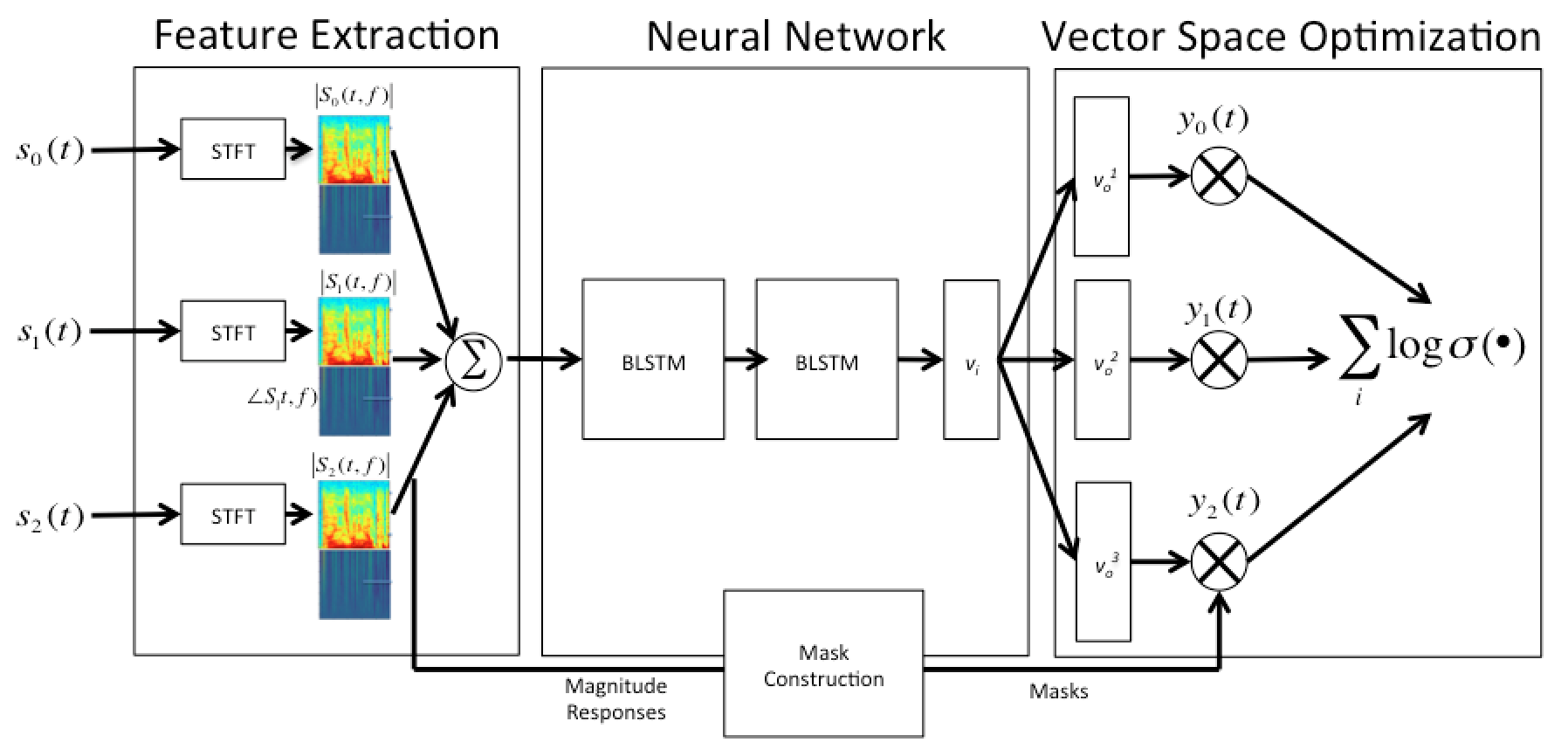}
\label{fig:training}
}
\subfigure[Inference]{
\includegraphics[width=0.85\textwidth]{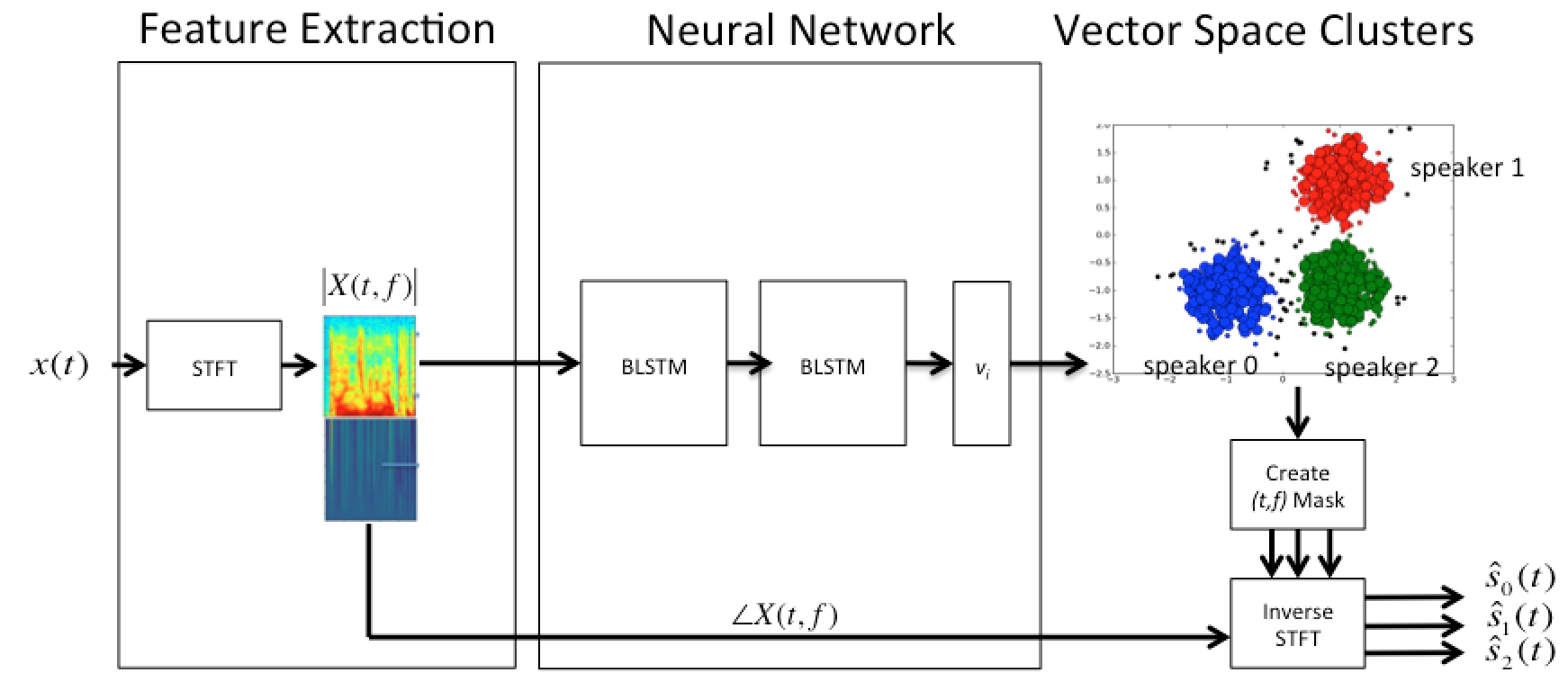}
\label{fig:inference}
} 
\caption{General architecture using source-contrastive estimation to optimize a vector space for speaker attributes} \label{fig:algorithm}
\end{figure*}

Additionally in Fig.~\ref{algorithm} during inference, we do not use the output vector space $V_o$. While true that computations are further reduced, the intention is that the out-of-set speaker set is allowed. In fact, even though we may train on mixes with fewer speakers, we can inference in situations where there are arbitrary numbers of speaker.


\section{Implementation}
\label{implementation}
The algorithm was implemented in Tensorflow, v1.1~\cite{tensorflow}. (We make extensive use of broadcasting operation, and earlier versions of the function {\small \verb"gather_nd"} will not pass gradients back.) The model architecture consists of two bidirectional long short term memory (BLSTM) layers $r_1, r_2$ of 600 units each. These are followed by a fully connected layer $d_1$ that maps the output of the second BLSTM layer to the input vector space, the final box in Fig.~\ref{fig:training}.  The BLSTM layers both use tanh nonlinearities, and the fully connected layer is linear.  For a batch of inputs $\mathbf{X}$, the output of the two BLSTM layers $r_1, r_2 \in \mathbb{R}^{B \times T \times 600}$. While the final layer of the neural network is technically a fully-connected linear layer, it is implemented as a $1D$ convolution over the $r_2$ output tensor with a filter $w\in\mathbb{R}^{1\times600\times F\cdot E}$.  The output of the convolution can then be reshaped to give the input vector space $\mathbf{V}_i\in\mathbb{R}^{B\times T \times F \times E}$.  This implementation allows the model to be run for arbitrary input $T$, which is useful at inference time.

For efficient evaluation of the cost function of Eq. \ref{eqn:cost} across batches, the speaker vectors for only speakers represented in each batch are assembled into a tensor $\mathbf{V}_o\in\mathbb{R}^{B \times M \times E}$.  The ordering of the $M$ speakers in $\mathbf{V}_o$ must match the ordering used in $\mathbf{Y}$, but is otherwise arbitrary.  To efficiently compute the dot products $V_i\cdot V_o$ in Eq. \ref{eqn:cost} with broadcasting, we expand the dimensions of the input vector space to $\mathbf{V}_i\in\mathbb{R}^{B \times T \times F \times 1 \times E}$ and the output vectors expands to $\mathbf{V}_o\in\mathbb{R}^{B\times 1\times 1 \times M \times E}$.  
This gives an output of the dot product operation as a tensor $\mathbf{D}\in\mathbb{R}^{B\times T\times F \times M}$, which is compatable with the labels $\mathbf{Y}$ and so they can be multiplied together elementwise to give the argument of the sigmoid in Eq. \ref{eqn:cost}.  The remaining portion of the cost function is easily evaluated.

Our batch size is $B=256$ during training. The input tensors have dimensions $\mathbf{X}\in\mathbb{R}^{B \times T \times F}$ and label tensors are $\mathbf{Y}\in\mathbb{R}^{B \times T \times F \times M}$, where $T=40$ is the length of total time steps per sample and $F=512$ are the number of frequency bins used.

\subsection{Preprocessing}

In all our experiments, signals were resampled to $10kHz$. To generate features, we scaled to zero mean and unity standard deviation. Prior to processing, we applied a preemphasis filter with coefficient $0.95$. The signals were then converted into short time Fourier transform (STFT) spectrograms. The parameters used were $0.0512 s$ ($512$ sample) Hanning windows of the input waveform with a $0.0256 s$ overlap, computing the Fourier transform of the signal in each window, and stacking the resultant spectrograms together to obtain a 2D T-F (Time-Frequency) representation of the signal.  The complex phases $\phi_{t,f}$ were saved separately for use in post separation processing. Separate spectrograms for the signal from each speaker ($S^{(i)}_{t,f}$ for $n\in\{1,2,...,C\}$) were computed for training and evaluation purposes, while the total spectrogram was computed by the elementwise sum $X_{t,f}=S^{(i)}_{t,f}+S^{(j)}_{t,f}$ for two speakers with IDs $n,m$.

The $X_{t,f}$ spectrograms were then passed through a square root nonlinearity and percent normalized.  This is similar to the procedure suggested in \cite{wang2014training}; however we obtained better results with a square root rather than a logarithmic nonlinearity.  This gives the input features for the model as 
\begin{equation}
\label{eqn:input_normalization}
\bar{X}_{t,f} = \frac{\sqrt{\left|X_{t,f}\right|} - \min{\sqrt{\left|X_{t,f}\right|}}}{\max{\sqrt{\left|X_{t,f}\right|}} - \min{\sqrt{\left|X_{t,f}\right|}}}
\end{equation}

Speaker labels $Y_{t,f}^{(c)}$ are assigned to each T-F bin by giving a value of $1$ to the signal which is loudest in at that time and frequency, and a value of $-1$ to all other signals.  
\begin{equation}
\label{eqn:labels}
Y_{t,f}^{(c)} = \left\{
    \begin{aligned}
    &1\text{ if }\text{argmax}\left(\left\{\left|S_{t,f}^{(1)}\right|,\left|S^{(2)}_{t,f}\right|,...,\left|S_{t,f}^{(M)}\right|\right\}\right)=c \\
    -&1\text{ if }\text{argmax}\left(\left\{\left|S_{t,f}^{(1)}\right|,\left|S_{t,f}^{(2)}\right|,...,\left|S_{t,f}^{(M)}\right|\right\}\right)\neq c
    \end{aligned} \right.
\end{equation}

\subsection{Reconstruction}

At inference time, a signal consisting of an unknown mixture of sources is preprocessed as described in the previous subsection, giving a complex T-F estimate of a single source signal, $\hat{S}_{t,f}$.  An input feature is generated as in Eq. \ref{eqn:input_normalization} and fed through the model to obtain the vectors $\mathbf{V}_i$.  A $K$-means clustering is then performed on the vectors in order to generate a labeling prediction $\mathbf{\hat{Y}}\in\mathbb{R}^{T\times F\times K}$ in which each T-F element is associated with a cluster label.  Here the element $\hat{Y}^{(k)}_{t,f}=1$ if the associated vector $V_{t,f}$ belongs to the $k^{th}$ cluster, and $\hat{Y}^{(k)}_{t,f}=-1$ otherwise as in Eq. \ref{eqn:labels}.  These labelings can then be used as masks to reconstruct a source $S^{(k)}_{t,f}$ from each of the $K$ clusters.  T-F representations of the inferred sources are calculated as the elementwise multiplication of the input spectrogram with the inferred labeling.
\begin{equation}
\hat{S}^{(k)}_{t,f} = X_{t,f}\odot \frac{1}{2} \left( \hat{Y}^{(k)}_{t,f} + 1 \right)
\end{equation}
The source spectrogram $\hat{S}^{(k)}_{t,f}$ is then converted (using the inverse STFT) into a source waveform and the preemphasis filtering is undone, completing the inference process.


All code involved in our contribution, those used to replicate research in~\cite{dong-permutation,hershey2016deep}, and evaluation code can be found at {\small \verb"http://github.com/lab41/magnolia"}.

\section{Experiments}
\label{results}
All models were trained on two speaker mixes generated from the {\small \verb"train-clean-100"} subset of the LibriSpeech ASR corpus \cite{librispeech}.  This subset consists of 100 hours of high quality read English speech from 251 speakers (125 females 126 males).  The training subset was split three ways: 80\% to train the model, 10\% to monitor the cost, and the final 10\% to compute the \emph{in-set} speakers via Signal-to-Distortion Ratio (SDR) improvement. Each of the splits contained unique readings not present in other splits from each of the 251 speakers. With the same model, the SDR for \emph{out-of-set} speakers were computed on Librispeech's {\small \verb"test-clean"} subset. Mixtures were generated by additively combining two different utterances from within each split to create mixed signals with known sources.  

In addition to sparse NMF~\cite{monaural-nmf} baselines, we compare against the latest state of the art to include the convolutional model (CNN) of PIT~\cite{dong-permutation} and Hershey et al.'s deep clustering~\cite{hershey2016deep}. The proposed method, source-contrastive estimation (SCE), is listed last in all tables.

\subsection{Vector embedding}

\begin{figure}[h]
\centering
\subfigure[Deep Clustering]{\includegraphics[width=0.2\textwidth]{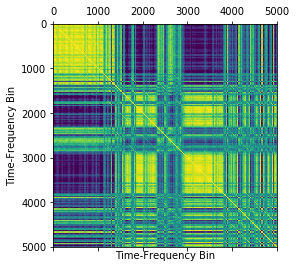}}
\subfigure[Proposed Algorithm]{\includegraphics[width=0.2\textwidth]{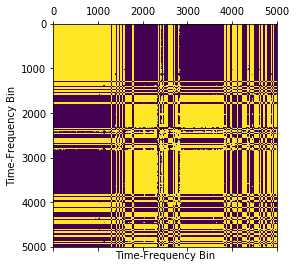}}
\caption{Comparison of vector-vector correlation vectors between deep clustering and the proposed vector space \emph{on the same color scale}. Note the in-between values for deep clustering versus the di-chromicity of the proposed algorithm. This is indicative of more confident assignments and better defined clusters.}
\label{fig:affinity}
\end{figure}
A comparison between the embedding vector space learned by deep clustering and the one learned by SCE can be seen in Fig. \ref{fig:affinity}.  While both models attempt to learn an embedding space such that the vectors $\mathbf{v}_i$ belonging to the same speaker are close to one another,  the optimization objectives of the two models are quite different.  Deep clustering optimizes this by taking the Frobenius norm of the difference between the inferred affinity matrix $VV^T$ and the known affinity matrix ($YY^T$ in their model) as the cost function.  Our model (SCE) does not optimize to the affinity matrix, but optimizes the vector space via~\eqref{eqn:cost} such that embedding vectors from the same speaker are part of their own cluster and far from embedding vectors belonging to other speakers.  This leads to a larger separation between clusters.  
\begin{figure}[h]
\centering
\includegraphics[width=0.49\textwidth]{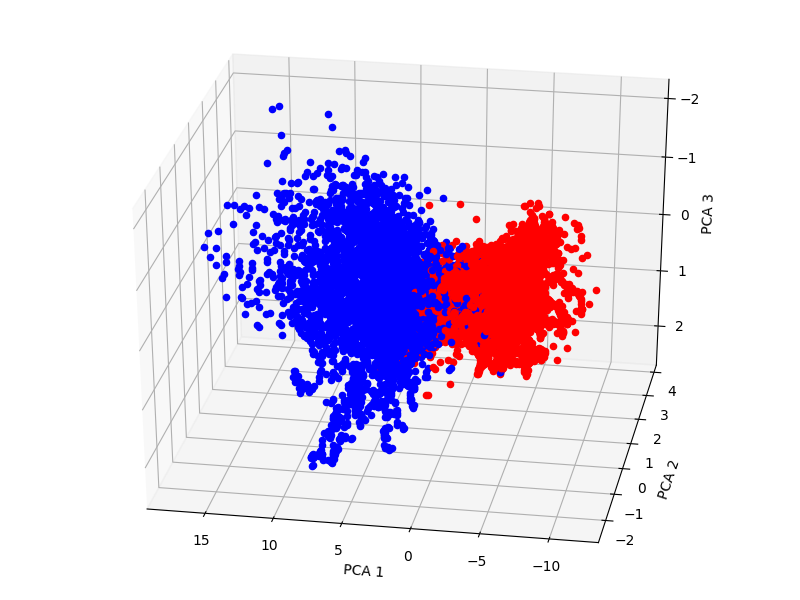}
\caption{Low-dimensional vector space projections of representations of two simultaneously speaking sources over five seconds.}
\label{fig:vector-space}
\end{figure}
A low-dimensional visualization of the vector embeddings for an example five second mixture can be seen in Fig. \ref{fig:vector-space}.  Each point represents one time-frequency bin color coded according to which speaker it corresponds to.  Vectors corresponding to different speakers lie in different regions of the high dimensional space.

\subsection{Signal-to-distortion ratio improvement}

Evaluation of the separation quality was performed on eight types of mixture sets.  As the problem of separating speech is significantly more difficult when both speakers are female or male, SDR improvement was calculated separately for female-female and male-male mixes.  In addition, female-male mixes were also evaluated, as well as mixes of uniformly random speakers from the testing splits.  The SDR improvement for mixes consisting of three randomly selected speakers was also evaluated in order to test the ability of models to generalize beyond the types of mixes used in training. 
\begin{table}
\begin{center}
\begin{tabular}{ |p{2.9cm}||p{.75cm} p{.75cm} p{.75cm} p{.75cm}|  }
 \hline
 & f+f & m+m & f+m & All\\
 \hline
 SNMF                        & 0.69 & 0.30 & 0.28 & 0.39 \\
 PIT-S-CNN~\cite{dong-permutation} & 4.29 & 4.05 & 7.21 & 5.69    \\
 Deep clustering~\cite{hershey2016deep} & 4.22 & \textbf{5.68} & 9.37 & 7.17 \\
 Proposed (SCE)              & \textbf{5.33} & 5.48 & \textbf{9.98} & \textbf{7.69}    \\
 \hline
\end{tabular}
\end{center}
\caption{SDR improvement for in set speaker mixtures}
\label{tbl:inset}
\end{table}
Since our model explicitly makes use of speaker identities during training, performance was measured separately on mixtures containing new utterances from the speakers in the training set.  Table \ref{tbl:inset} shows the evaluated SDR improvement for each of the four mix types for speakers used in training. For both female-female and female-male mixtures, our model outperforms  deep clustering  \cite{hershey2016deep} and maintains comparable performance on male-male mixtures.  
\begin{table}[h!]
\begin{center}
\begin{tabular}{ |p{2.5cm}||p{0.75cm} p{0.75cm} p{0.75cm} p{0.75cm}|  }
 \hline
 & f+f & m+m & f+m & All\\
 \hline
 PIT-S-CNN       & 4.23    & 3.88 & 7.10 & 5.58\\
 Deep clustering & 4.01 & \textbf{4.94} & 8.84 & 6.66\\
 Proposed (SCE)   & \textbf{4.39} & 3.98 & \textbf{10.09} & \textbf{7.14} \\
 \hline
\end{tabular}
\end{center}
\caption{SDR improvement for out of set speaker mixtures}
\label{tbl:outset}
\end{table}
SDR improvement on mixtures of speakers that were not in the training set (from the test-clean subset of the LibriSpeech corpus) can be seen in Table \ref{tbl:outset}.  As the application of NMF to this type of mixture is not well defined, this model was not evaluated on these mixtures.  While the SDR improvement is somewhat lower than for the in set mixtures, our model still favorably compares to the other related models, outperforming both deep clustering and the PIT-S-CNN model on female-female and female-male mixtures.    
\begin{table}[h!]
\begin{center}
\begin{tabular}{ |p{2.5cm}||p{2cm} p{2cm}|  }
 \hline
 & In-set all & Out-of-set all\\
 \hline
 Deep clustering & \textbf{5.01}    & 4.33 \\
 Proposed (SCE)  & 4.98    & \textbf{4.37} \\
 \hline
\end{tabular}
\end{center}
\caption{SDR improvement for mixtures of three speakers}
\label{tbl:threemix}
\end{table}
Despite being trained only on mixes of two speakers, both deep clustering and our model have some ability to separate mixtures of more than two speakers.  Performance on randomly selected mixtures of three speakers is given in Table \ref{tbl:threemix}.  While both models give an improvement in SDR, our model performs somewhat better on out-of-set separation. 

\subsection{Computational complexity}
\begin{table}[t]
\begin{center}
\begin{tabular}{ |p{2.5cm}||p{2.5cm}|  }
 \hline
 & Batch time ($s$)\\
 \hline
 Deep clustering   & 0.45 \\
 PIT-S-CNN     &   0.39 \\
 Proposed (SCE)    &   \textbf{0.21}  \\
 \hline
\end{tabular}
\end{center}
\caption{Time to train on a batch of size $B=256$}
\label{tbl:batchtime}
\end{table}
During training, the runtime of both DC and PIT are a concern. On our NVIDIA Titan X card, DC and PIT's CNN-based model takes as much as twice the time to do a single training update on a batch of 256 examples as our model takes (see table \ref{tbl:batchtime}), averaged over thousands of training iterations.


\section{Conclusions}
\label{conclusions}
The proposed method uses a recurrent neural network to project speaker characteristics into an embedding space to perform source separation using contrastive estimation. Doing so achieves a 7.3dB SDR gain compared to the SNMF baseline and 0.48-2 dB SDR gains over other deep learning approaches on in-set speakers and 0.48 to 1.56 dB SDR gain over out-of-set speakers. Our approach saves computational time due to a less burdensome sampling-based technique. Future work 
involves alternative feature representations not limited to cepstral-based features. As the proposed work replicates BLSTMs for comparisons-sake, we can also explore other RNN architectures and layers (GRU's) to reduce computational load.


\bibliographystyle{ieeetr}
\bibliography{main}

\begin{thebibliography}{10}

\bibitem{bss-beamforming}
J.~Sanz-Robinson, L.~Huang, T.~Moy, W.~Rieutort-Louis, Y.~Hu, S.~Wagner, J.~C.
  Sturm, and N.~Verma, ``Robust blind source separation in a reverberant room
  based on beamforming with a large-aperture microphone array.,'' in {\em
  ICASSP}, pp.~440--444, IEEE, 2016.

\bibitem{monaural-ica}
L.~K. Hansen and K.~B. Petersen, ``Monaural {ICA} of white noise mixtures is
  hard,'' in {\em Proceedings of {ICA'}2003 Fourth Int. Symp.. on Independent
  Component Analysis and Blind Signal Separation, Nara Japan, April 4,},
  pp.~815--820, 2003.

\bibitem{monaural-nmf}
T.~Virtanen, ``Monaural sound source separation by nonnegative matrix
  factorization with temporal continuity and sparseness criteria,'' {\em IEEE
  Transactions on Audio, Speech, and Language Processing}, vol.~15,
  pp.~1066--1074, March 2007.

\bibitem{isik16}
Y.~Isik, J.~L. Roux, Z.~Chen, S.~Watanabe, and J.~R. Hershey, ``Single-channel
  multi-speaker separation using deep clustering,'' {\em CoRR,
  http://arxiv.org/abs/1607.02173}, vol.~abs/1607.02173, 2016.

\bibitem{KandT}
Z.~Koldovský and P.~Tichavský, ``Time-domain blind separation of audio
  sources on the basis of a complete ica decomposition of an observation
  space,'' {\em IEEE Trans. on Speech, Audio and Language Processing}, vol.~19,
  pp.~406--416, February 2011.

\bibitem{schmidt-nmf}
M.~N. Schmidt, ``Speech separation using non-negative features and sparse
  non-negative matrix factorization,'' in {\em Computer Speech and Language,
  2008, submitted. [Online]. Available:
  http://www.imm.dtu.dk/pubdb/p.php?5377}.

\bibitem{dong-permutation}
D.~Yu, M.~Kolb{\ae}k, Z.~Tan, and J.~Jensen, ``Permutation invariant training
  of deep models for speaker-independent multi-talker speech separation,'' {\em
  CoRR}, vol.~abs/1607.00325, 2016.

\bibitem{deepkaraoke}
A.~J.~R. Simpson, G.~Roma, and M.~D. Plumbley, ``Deep karaoke: Extracting
  vocals from musical mixtures using a convolutional deep neural network,''
  {\em CoRR}, vol.~abs/1504.04658, 2015.

\bibitem{hershey2016deep}
J.~R. Hershey, Z.~Chen, J.~Le~Roux, and S.~Watanabe, ``Deep clustering:
  Discriminative embeddings for segmentation and separation,'' in {\em
  Acoustics, Speech and Signal Processing (ICASSP), 2016 IEEE International
  Conference on}, pp.~31--35, IEEE, 2016.

\bibitem{pit-ref-1}
C.~Weng, D.~Yu, M.~L. Seltzer, and J.~Droppo, ``Deep neural networks for
  single-channel multi-talker speech recognition,'' {\em IEEE/ACM Transactions
  on Audio, Speech, and Language Processing}, vol.~23, pp.~1670--1679, Oct
  2015.

\bibitem{mikolov}
T.~Mikolov, I.~Sutskever, K.~Chen, G.~S. Corrado, and J.~Dean, ``Distributed
  representations of words and phrases and their compositionality,'' in {\em
  Advances in Neural Information Processing Systems 26} (C.~J.~C. Burges,
  L.~Bottou, M.~Welling, Z.~Ghahramani, and K.~Q. Weinberger, eds.),
  pp.~3111--3119, Curran Associates, Inc., 2013.

\bibitem{attractor}
Z.~Chen, Y.~Luo, and N.~Mesgarani, ``Deep attractor network for
  single-microphone speaker separation,'' {\em CoRR}, vol.~abs/1611.08930,
  2016.

\bibitem{tensorflow}
M.~Abadi, A.~Agarwal, P.~Barham, E.~Brevdo, Z.~Chen, C.~Citro, G.~S. Corrado,
  A.~Davis, J.~Dean, M.~Devin, S.~Ghemawat, I.~Goodfellow, A.~Harp, G.~Irving,
  M.~Isard, Y.~Jia, R.~Jozefowicz, L.~Kaiser, M.~Kudlur, J.~Levenberg,
  D.~Man\'{e}, R.~Monga, S.~Moore, D.~Murray, C.~Olah, M.~Schuster, J.~Shlens,
  B.~Steiner, I.~Sutskever, K.~Talwar, P.~Tucker, V.~Vanhoucke, V.~Vasudevan,
  F.~Vi\'{e}gas, O.~Vinyals, P.~Warden, M.~Wattenberg, M.~Wicke, Y.~Yu, and
  X.~Zheng, ``{TensorFlow}: Large-scale machine learning on heterogeneous
  systems,'' 2015.
\newblock Software available from tensorflow.org.

\bibitem{wang2014training}
Y.~Wang, A.~Narayanan, and D.~Wang, ``On training targets for supervised speech
  separation,'' {\em IEEE/ACM Transactions on Audio, Speech and Language
  Processing (TASLP)}, vol.~22, no.~12, pp.~1849--1858, 2014.

\bibitem{librispeech}
V.~Panayotov, G.~Chen, D.~Povey, and S.~Khudanpur, ``Librispeech: an asr corpus
  based on public domain audio books,'' in {\em Acoustics, Speech and Signal
  Processing (ICASSP), 2015 IEEE International Conference on}, pp.~5206--5210,
  IEEE, 2015.

\end{thebibliography}

\end{document}